# A polynomial-time algorithm for the ground state of one-dimensional gapped Hamiltonians

Yichen Huang (黄溢辰)[*]
University of California, Berkeley, Berkeley, California 94720, USA
California Institute of Technology, Pasadena, California 91125, USA
ychuang@caltech.edu

October 26, 2015

**Abstract**

A (deterministic) polynomial-time algorithm is proposed for approximating the ground state of (general) one-dimensional gapped Hamiltonians. Let $\epsilon, n, \eta$ be the energy gap, the system size, and the desired precision, respectively. Neglecting $\epsilon$-dependent subpolynomial (in $n$) and constant factors, the running time of the algorithm is $n^{O(1)}$ for $\eta = n^{-O(1)}$.

## 1 Introduction and background

Computing the ground state (energy) in quantum many-body systems with local interactions is a fundamental problem in condensed matter physics and a canonical problem in Hamiltonian complexity [21, 9]. Intuitively, this problem is likely intractable because the dimension of the Hilbert space of the system grows exponentially with the system size. Indeed, computing the ground-state energy of one-dimensional (1D) Hamiltonians is QMA-complete [3, 10]. Therefore, (assuming QMA≠NP) ground states of 1D Hamiltonians do not in general have classical representations from which physical properties can be efficiently extracted. It should be emphasized that the local Hamiltonians constructed in all known proofs of the QMA-hardness are gapless. Indeed, the ground state of 1D gapped Hamiltonians can be [12, 5, 15] efficiently represented as a matrix product state (MPS) [22, 8], a data structure that allows efficient computation of physical observables. Thus, the 1D gapped local Hamiltonian problem is in NP.

In practice, the density matrix renormalization group (DMRG) algorithm [27, 28] is highly successful in 1D gapped systems, and moderately successful in a large class of 1D gapless systems. It is the leading numerical method in 1D [23, 24] and is now running on the computers of condensed matter physicists everywhere on earth. Despite its remarkable popularity, DMRG is still a heuristic local search algorithm over MPS: It can get stuck in a local minimum and there is no guarantee that it always converges in polynomial time. The worst-case performance of DMRG-like algorithms has been a long-standing problem for more than two decades. Is there a variant of DMRG that provably finds the ground state of 1D gapped Hamiltonians in polynomial time, or is the 1D gapped local Hamiltonian problem in P? We are in a situation reminiscent of the practical success of the simplex algorithm for linear programming before the advent of the ellipsoid and interior-point methods.

A lot of progress has been made. Without assuming an energy gap, the ground-state energy of 1D commuting Hamiltonians can be computed efficiently using dynamic programming [25, 1].

---

[*]Supported by NSF DMR-1206515 and the Institute for Quantum Information and Matter.

This algorithm also provably finds the ground state of 1D gapped Hamiltonians in subexponential time [5]. In a recent remarkable paper, Landau, Vazirani, and Vidick [16] proposed a randomized polynomial-time algorithm for approximating the ground state of "almost frustration-free" 1D gapped Hamiltonians (in general 1D gapped systems its running time may be exponential in the system size). However, settling the complexity of the 1D gapped local Hamiltonian problem is still highly desirable because (a) a generic Hamiltonian is not almost frustration-free; (b) frustration-free Hamiltonians are expected to be easier to solve, e.g., they do not suffer from the so-called sign problem in quantum Monte Carlo simulations.

In the present paper, a (deterministic) polynomial-time algorithm is proposed for approximating the ground state of (general) 1D gapped Hamiltonians, i.e., we prove that the 1D gapped local Hamiltonian problem is in P. Thus, this algorithm has a much broader scope than the algorithm in [16]. As an immediate corollary, adiabatic quantum computation with a final 1D gapped Hamiltonian can be efficiently simulated classically and (assuming BQP≠P) is therefore not universal, improving Hastings' result [13] that adiabatic quantum computation with a path of 1D gapped Hamiltonians allows efficient classical simulation.

## 2 Main results

Suppose we are working with a chain of $n$ spins (qu$d$its), and the local dimension $d = \Theta(1)$ of each spin is an absolute constant. Let $\mathcal{H}_i = \mathbf{C}^d$ be the Hilbert space of the spin $i$; define $\mathcal{H}_{[i,j]} = \bigotimes_{k=i}^{j} \mathcal{H}_k$ as the Hilbert space of the spins with indices in the interval $[i,j]$ and $\mathcal{H} = \mathcal{H}_{[1,n]}$ as the Hilbert space of the system. Since the standard bra-ket notation can be cumbersome, in most but not all cases quantum states and their inner products are simply denoted by $\psi, \phi \ldots$ and $\langle \psi, \phi \rangle$, respectively, cf. $\||\psi\rangle - |\phi\rangle\|$ versus $\|\psi - \phi\|$. All states are normalized unless otherwise stated.

Let $H = \sum_{i=1}^{n-1} H_i$ be a 1D Hamiltonian, where $H_i$ acts on the spins $i$ and $i+1$ (nearest-neighbor interaction). Assume without loss of generality that the ground-state energy of each $H_i$ is zero, and $H_i \leq 1$. Let $\epsilon_0$ denote the ground-state energy of $H$, which is a measure of how frustrated $H$ is: $H$ is frustration-free if $\epsilon_0 = 0$, and $H$ is "almost frustration-free" if $\epsilon_0 = O(1)$. Suppose $H$ has a unique ground state $\Psi_0$ (Section 7 extends all results to the case of degenerate ground states), and there is a constant gap $\epsilon$ between the energies of the ground state and the first excited state. It is easy to see $\epsilon \leq 1$ (Lemma 11). The goal is to find an efficient MPS approximation to the ground state of $H$.

**Definition 1** (matrix product state (MPS) [22, 8]). Let $\{|j_i\rangle\}_{j_i=1}^d$ be the computational basis of $\mathcal{H}_i$ and $\{D_i\}_{i=0}^n$ with $D_0 = D_n = 1$ be a sequence of positive integers. An MPS $\Psi$ takes the form

$$\Psi = \sum_{j_1,j_2,\ldots,j_n=1}^{d} A_{j_1}^{[1]} A_{j_2}^{[2]} \cdots A_{j_n}^{[n]} |j_1 j_2 \cdots j_n\rangle, \qquad (1)$$

where $A_{j_i}^{[i]}$ is a matrix of size $D_{i-1} \times D_i$. Define $D = \max\{D_i\}_{i=0}^n$ as the bond dimension of the MPS $\Psi$.

Clearly, an MPS representation is efficient if its bond dimension is (at most) a polynomial in $n$. The existence of an efficient MPS approximation to the ground state of 1D gapped Hamiltonians is a by-product of the proof of the area law for entanglement [12, 5, 15, 26]. Let $\tilde{O}(x) := O(x \operatorname{polylog} x)$ hide a polylogarithmic factor, and suppose the desired precision $\eta = n^{-O(1)}$ is lower bounded by an inverse polynomial in $n$.

**Lemma 1** ([5, 15]). *There exists an MPS $\Psi$ of bond dimension $D = 2^{\tilde O(1/\epsilon + \epsilon^{-1/4} \log^{3/4}(n/\eta))}$ such that $|\langle \Psi, \Psi_0\rangle| \geq 1 - \eta$.*

As a corollary, the ground state of 1D gapped Hamiltonians can be computed in subexponential time using dynamic programming.

**Lemma 2** ([5, 25, 1]). *In 1D gapped systems there is a $2^{2^{\tilde O(\epsilon^{-1/4} \log^{3/4} n)}}$-time algorithm that outputs an MPS $\Psi$ such that $|\langle \Psi, \Psi_0\rangle| \geq 1 - \eta$.*

In a recent remarkable paper, Landau, Vazirani, and Vidick [16] proposed a randomized polynomial-time algorithm for approximating the ground state of almost frustration-free 1D gapped Hamiltonians ($\epsilon_0 = O(1)$).

**Lemma 3** ([16]). *In 1D gapped systems there is a randomized $n^{2^{\tilde O(1/\epsilon)} + O(\epsilon_0/\epsilon)}$-time algorithm that outputs an MPS $\Psi$ such that $|\langle \Psi, \Psi_0\rangle| \geq 1 - \eta$ with probability at least $1 - 1/\operatorname{poly} n$.*

*Remark.* In general 1D gapped systems its running time may be exponential in $n$ if $\epsilon_0 = \Theta(n)$.

Some $\epsilon$-dependent subpolynomial (e.g., $2^{\tilde O(\epsilon^{-1/4} \log^{3/4} n)}$) and constant (e.g., $2^{2^{\tilde O(1/\epsilon)}}$) factors will appear below. If not dominant (e.g., accompanied with $\operatorname{poly} n$), depending on the context they may be neglected or kept for simplicity or clarity, respectively. The main result of the present paper is

**Theorem 1.** *In (general) 1D gapped systems there is a (deterministic) $n^{O(1)}$-time algorithm that outputs an MPS $\Psi$ such that $|\langle \Psi, \Psi_0\rangle| \geq 1 - \eta$.*

This algorithm can be slightly improved in frustration-free 1D gapped systems.

**Theorem 2.** *In frustration-free 1D gapped systems there is a (deterministic) $n^{\omega + 2 + o(1)}$-time algorithm that outputs an MPS $\Psi$ such that $\langle \Psi, H\Psi\rangle \leq \eta\epsilon$ for $\eta = n^{-o(1)}$, where $\omega < 2.3729$ [29, 17] is the exponent of matrix multiplication.*

*Remark.* The exponent $\omega + 2 + o(1)$ is probably not tight, and we strongly believe that it can be improved to $\omega + 1 + o(1)$ without significant difficulties. It may even finally be improved to $1 + o(1)$.

Hastings [13] proved that adiabatic quantum computation with a path of 1D gapped Hamiltonians allows efficient classical simulation and (assuming BQP$\neq$P) is therefore not universal. Suppose $H(t)$ with $0 \leq t \leq t_{max} \leq \operatorname{poly} n$ is a "smooth" path of 1D Hamiltonians, where the ground state of $H(0)$ is a simple product state. Let $\Psi_0(t), \epsilon(t)$ be the ground state and the energy gap of $H(t)$, respectively, and $\eta = n^{-O(1)}$.

**Lemma 4** ([13]). *Suppose $H(t)$ has a constant energy gap for $0 \leq t \leq t_{max}$. Then there is an $n^{O(1/\min_{0\leq t\leq t_{max}}\{\epsilon(t)\})}$-time algorithm that outputs an MPS $\Psi$ such that $|\langle \Psi, \Psi_0(t_{max})\rangle| \geq 1 - \eta$.*

As an immediate corollary of Theorem 1, adiabatic quantum computation with a final 1D gapped Hamiltonian can be efficiently simulated classically. Suppose $H(t_{max})$ has a constant energy gap. Then the algorithm in Theorem 1 outputs an MPS $\Psi$ in time $n^{O(1)}$ such that $|\langle \Psi, \Psi_0(t_{max})\rangle| \geq 1 - \eta$.

## 3  Overview

The outline of the algorithm in Theorem 1 is similar to that in Lemma 3. We begin by defining the notion of "support set" (known as "viable set" in [16]).

Table 1: Evolution of the parameters in each iteration. The asterisks mark the parameter that is reduced at every step.

|                       | $i$   | $s$        | $b$           | $\delta$ | $\Delta$       |
|----------------------:|:-----:|:----------:|:-------------:|:--------:|:--------------:|
| start                 | $i-1$ | $p_1 p_3$  | $p_2 p_3$     | n/a      | $c\epsilon^6$  |
| **extension**         | $i$   | $dp_1 p_3$ | $p_2 p_3$     | n/a      | $c\epsilon^6$  |
| **cardinality reduction** | $i$ | $p_1{}^*$ | $dp_1 p_2 p_3^2$ | 1/1000 | 1/1000 |
| **bond truncation**   | $i$   | $p_1$      | $p_2{}^*$     | 1/20     | n/a            |
| **error reduction**   | $i$   | $p_1 p_3$  | $p_2 p_3$     | n/a      | $c\epsilon^{6}*$ |

**Definition 2** (support set). $S \subseteq \mathcal{H}_{[1,i]}$ is an $(i, s, b, \delta$ or $\Delta)$-support set if there exists a state $\psi \in \mathcal{H}$ (called a witness for $S$) such that
(i) the reduced density matrix of $\psi$ on $\mathcal{H}_{[1,i]}$ is supported on span $S$;
(ii) $|S| \leq s$;
(iii) all elements in $S$ are MPS of bond dimension at most $b$;
(iv) $|\langle \psi, \Psi_0 \rangle| \geq 1 - \delta$ or $\langle \psi, H\psi \rangle \leq \epsilon_0 + \Delta\epsilon$ (depending on the context either $\delta$ or $\Delta$ is used as the precision parameter).

*Remark.* Lemma 5 implies that an $(i, s, b, \Delta = \eta)$-support set is also an $(i, s, b, \delta = \eta)$-support set.

The algorithm in Theorem 1 iteratively constructs an $(i, p_1 p_3, p_2 p_3, \Delta = c\epsilon^6)$-support set $S_i$ for $i = 1, 2, \ldots, n$, where $p_1, p_2, p_3$ are (upper bounded by) $i$-independent and $\epsilon$-dependent polynomials in $n$, and $c$ is a sufficiently small absolute constant. After the last iteration, we obtain an MPS approximation to the ground state $\Psi_0$ of $H$ from the last support set $S_n$ by minimizing the energy over the subspace span $S_n$. By definition, the solution has precision $\Delta = c\epsilon^6$. The minimization can be efficiently formulated as a convex program. The construction of this convex program requires contracting the MPS in $S_n$. The size of the convex program is determined by $\dim \operatorname{span} S_n$, which is a polynomial in $n$. Finally, we improve the precision of the solution to $\delta = \eta$ in time $n^{O(1)}$.

Each iteration consists of four steps: **extension**, **cardinality reduction**, **bond truncation**, and **error reduction**. Table 1 summarizes the evolution of the parameters $s, b, \delta$ or $\Delta$ in each iteration of the algorithm in Theorem 1.

We briefly recall the algorithm in Lemma 3. This algorithm only uses $\delta$ as the precision parameter, which is reduced to $O(\epsilon^2/n)$ at the end of the $i$th iteration for $i = 1, 2, \ldots, n-1$ (the $n$th iteration is slightly different). The analysis in [16] gives

$$p_1 = n2^{\tilde{O}(1/\epsilon)}, \ p_2 = n2^{\tilde{O}(\epsilon^{-1/4}\log^{3/4} n)}, \ p_3 = n^{O(1/\epsilon + \epsilon_0/\epsilon)} \tag{2}$$

in general 1D gapped systems. The running time of the algorithm is a polynomial in $p_1, p_2, p_3$, and is dominated by $p_1$ in almost frustration-free 1D gapped systems ($\epsilon_0 = O(1)$). Specifically, **extension** is trivial. **Cardinality reduction** is based on the technique of dynamic programming for MPS [25, 1] and requires solving convex programs. Its analysis determines $p_1$. **Bond truncation** is simple and straightforward. Its analysis determines $p_2$. **Error reduction** is the only step that involves randomness (it succeeds with probability at least $1 - 1/\operatorname{poly} n$). Its analysis determines $p_3$, which depends exponentially on $\epsilon_0$.

In the present paper, we significantly improve the analysis of **cardinality reduction** using perturbation theory (the truncation lemma [5, 15]) so that $p_1 = 2^{2^{\tilde{O}(1/\epsilon)}}$ never dominates the running time of the algorithm. More importantly, we redesign **error reduction**. In general 1D gapped systems we use the Fourier transform and the Lieb-Robinson bound [18, 19, 14] (but not randomness) so that $p_3 = n^{O(1)}$. In frustration-free 1D gapped systems we use the detectability lemma [2, 4] so that $p_3 = 2^{\tilde{O}(1/\epsilon)}$; thus, all convex programs are of subpolynomial size. The other two (easy) steps (**extension** and **bond truncation**) remain unchanged.

## 4 Preliminaries

**Lemma 5.** $\langle \psi, H\psi \rangle \leq \epsilon_0 + \eta\epsilon$ *implies* $|\langle \psi, \Psi_0 \rangle| \geq |\langle \psi, \Psi_0 \rangle|^2 \geq 1 - \eta$.

*Proof.* The state $\psi$ can be decomposed as $\psi = c_0 \Psi_0 + c_1 \Psi_1$, where $\Psi_1 \perp \Psi_0$ and $\langle \Psi_1, H\Psi_1 \rangle \geq \epsilon_0 + \epsilon$.

$$\epsilon_0 + \eta\epsilon \geq \langle \psi, H\psi \rangle \geq |c_0|^2 \epsilon_0 + |c_1|^2 (\epsilon_0 + \epsilon) = \epsilon_0 + |c_1|^2 \epsilon \Rightarrow |c_0|^2 = 1 - |c_1|^2 \geq 1 - \eta. \tag{3}$$

$\square$

**Lemma 6.** $|\langle \psi, \phi_1 \rangle| \geq 1 - \eta_1$ *and* $|\langle \psi, \phi_2 \rangle| \geq 1 - \eta_2$ *imply* $|\langle \phi_1, \phi_2 \rangle| \geq 1 - 2(\eta_1 + \eta_2)$.

*Proof.* Let $\theta_1$ be the angle between $\psi$ and $\phi_1$, and $\theta_2$ be the angle between $\psi$ and $\phi_2$.

$$\begin{aligned}|\langle \phi_1, \phi_2 \rangle| &\geq \cos(\theta_1 + \theta_2) = \cos\theta_1 \cos\theta_2 - \sqrt{(1 - \cos^2\theta_1)(1 - \cos^2\theta_2)} \\ &\geq (1 - \eta_1)(1 - \eta_2) - \sqrt{(2\eta_1 - \eta_1^2)(2\eta_2 - \eta_2^2)} \geq 1 - 2(\eta_1 + \eta_2).\end{aligned} \tag{4}$$

$\square$

**Lemma 7.** $|\langle \psi, \phi \rangle| \geq 1 - \eta$ *implies* $|\langle \psi, \hat{O}\psi \rangle - \langle \phi, \hat{O}\phi \rangle| \leq 2\sqrt{2\eta}$ *for any operator* $\hat{O}$ *with* $\|\hat{O}\| \leq 1$.

*Proof.* Assume without loss of generality that $\langle \psi, \phi \rangle$ is a positive real number.

$$\begin{aligned}\|\psi - \phi\|^2 &= 2 - 2\langle \psi, \phi \rangle \leq 2\eta \text{ and } \langle \psi, \hat{O}\psi \rangle - \langle \phi, \hat{O}\phi \rangle = \langle \psi - \phi, \hat{O}\psi \rangle + \langle \phi, \hat{O}(\psi - \phi) \rangle \\ &\Rightarrow |\langle \psi, \hat{O}\psi \rangle - \langle \phi, \hat{O}\phi \rangle| \leq \|\psi - \phi\| \cdot \|\hat{O}\| \cdot \|\psi\| + \|\phi\| \cdot \|\hat{O}\| \cdot \|\psi - \phi\| \leq 2\sqrt{2\eta}.\end{aligned} \tag{5}$$

$\square$

**Definition 3** (truncation). Let $\psi = \sum_{j \geq 1} \lambda_j l_j \otimes r_j$ be the Schmidt decomposition of a state $\psi \in \mathcal{H}$ across the cut $i|i+1$, where the Schmidt coefficients are in descending order: $\lambda_1 \geq \lambda_2 \geq \cdots > 0$. Define $\operatorname{trunc}_D \psi = \sum_{j=1}^{D} \lambda_j l_j \otimes r_j$.

**Lemma 8** (Eckart-Young theorem [7]). *The state* $\psi' = \operatorname{trunc}_D \psi / \|\operatorname{trunc}_D \psi\|$ *satisfies* $\langle \psi', \psi \rangle \geq |\langle \phi, \psi \rangle|$ *for any state* $\phi \in \mathcal{H}$ *of Schmidt rank* $D$ *(across the cut* $i|i+1$).

**Lemma 9** ([16]). *Suppose* $\phi \in \mathcal{H}$ *is a state of Schmidt rank* $D$ *(across the cut* $i|i+1$).

$$|\langle \operatorname{trunc}_{D/\eta} \psi, \phi \rangle| \geq |\langle \psi, \phi \rangle| - \eta, \ \forall \eta > 0, \psi \in \mathcal{H}. \tag{6}$$

*Remark.* This is a simple corollary of Lemma 8.

**Lemma 10.** $\langle \Psi'_0, \Psi_0 \rangle \geq 1 - \eta$ *for* $\Psi'_0 = \operatorname{trunc}_{B_\eta} \Psi_0 / \|\operatorname{trunc}_{B_\eta} \Psi_0\|$, *where* $B_\eta = 2^{\tilde{O}(1/\epsilon + \epsilon^{-1/4} \log^{3/4}(1/\eta))}$.

*Proof.* As a by-product of the proof of the area law for entanglement [15], there exists a state $\phi \in \mathcal{H}$ of Schmidt rank $B_\eta$ such that $\langle \phi, \Psi_0 \rangle \geq 1 - \eta$. Then, Lemma 10 follows from Lemma 8. $\square$

**Lemma 11.** $\epsilon \leq 1$.

*Proof.* It suffices to find two orthogonal states with energies at most $\epsilon_0 + 1$. Let $\{|j_1\rangle\}_{j_1=1}^d$ be the computational basis of $\mathcal{H}_1$, and $\psi \in \mathcal{H}_{[2,n]}$ be the ground state of $\sum_{i=2}^{n-1} H_i$. For any $|j_1\rangle$,

$$\langle j_1\psi, Hj_1\psi\rangle = \langle j_1\psi, H_1 j_1\psi\rangle + \left\langle \psi, \sum_{i=2}^{n-1} H_i\psi \right\rangle \le \langle \Psi_0, H_1\Psi_0\rangle + 1 + \left\langle \Psi_0, \sum_{i=2}^{n-1} H_i\Psi_0 \right\rangle = \epsilon_0 + 1. \quad (7)$$

□

**Lemma 12.** $\langle\psi, H\psi\rangle \le \epsilon_0 + \eta\epsilon$ and $\eta \le 1/10$ *imply* $\langle\psi', H\psi'\rangle \le \epsilon_0 + 25\sqrt{\eta}$ *for* $\psi' = \text{trunc}_{B_\eta}\psi/\|\text{trunc}_{B_\eta}\psi\|$.

*Proof.* Lemma 5 implies $|\langle\psi, \Psi_0\rangle| \ge 1 - \eta$. Since $|\langle\Psi_0', \Psi_0\rangle| \ge 1 - \eta$ for $\Psi_0' = \text{trunc}_{B_\eta}\Psi_0/\|\text{trunc}_{B_\eta}\Psi_0\|$ (Lemma 10), Lemma 6 implies $|\langle\psi, \Psi_0'\rangle| \ge 1 - 4\eta$. Let $\phi' = \text{trunc}_{B_\eta}\psi$ and $\psi = \phi' + \phi$. Lemma 8 implies $\|\phi'\| = \langle\psi, \phi'\rangle/\|\phi'\| \ge 1 - 4\eta$. Hence, $\|\phi\|^2 = 1 - \|\phi'\|^2 \le 8\eta$. Since $\eta \le 1/10$,

$$\begin{aligned}
&\langle\psi, (H - H_i)\psi\rangle = \langle\phi', (H - H_i)\phi'\rangle + \langle\phi, (H - H_i)\phi\rangle \\
&\Rightarrow \langle\phi', H\phi'\rangle = \langle\psi, H\psi\rangle - \langle\phi, H\phi\rangle + \langle\phi, H_i\phi\rangle - \langle\phi', H_i\phi'\rangle - \langle\phi, H_i\psi\rangle \\
&\le \epsilon_0 + \eta\epsilon - \epsilon_0\|\phi\|^2 + \|\phi\|^2 + 2\|\phi\| \le \epsilon_0\|\phi'\|^2 + \eta\epsilon + 8\eta + 4\sqrt{2\eta} \\
&\Rightarrow \langle\psi', H\psi'\rangle \le \epsilon_0 + (\eta\epsilon + 8\eta + 4\sqrt{2\eta})/(1 - 4\eta)^2 \le \epsilon_0 + 25\sqrt{\eta}.
\end{aligned} \quad (8)$$

□

**Definition 4** (matrix product operator (MPO))**.** Let $\{\hat{O}_{j_i}\}_{j_i=1}^{d^2}$ be a basis of the space of operators on $\mathcal{H}_i$ and $\{D_i\}_{i=0}^n$ with $D_0 = D_n = 1$ be a sequence of positive integers. As the operator analog of MPS, an MPO $K$ takes the form

$$K = \sum_{j_1, j_2, \ldots, j_n = 1}^{d^2} \left( A_{j_1}^{[1]} A_{j_2}^{[2]} \cdots A_{j_n}^{[n]} \right) \hat{O}_{j_1} \otimes \hat{O}_{j_2} \otimes \cdots \otimes \hat{O}_{j_n}, \quad (9)$$

where $A_{j_i}^{[i]}$ is a matrix of size $D_{i-1} \times D_i$. Define $D = \max\{D_i\}_{i=0}^n$ as the bond dimension of the MPO $K$.

## 5 Algorithm and analysis

In the $i$th iteration the algorithm in Theorem 1 constructs an $(i, p_1p_3, p_2p_3, \Delta = c\epsilon^6)$-support set $S_i$ from an $(i-1, p_1p_3, p_2p_3, \Delta = c\epsilon^6)$-support set $S_{i-1}$ returned in the $(i-1)$th iteration.

### 5.1 Extension

**Extension** is trivial. Let $\{|j_i\rangle\}_{j_i=1}^d$ be the computational basis of $\mathcal{H}_i$. It is easy to see that $S_i^{(1)} := \{\psi|j_i\rangle : \forall \psi \in S_{i-1}, \ j_i = 1, 2, \ldots, d\}$ is an $(i, dp_1p_3, p_2p_3, \Delta = c\epsilon^6)$-support set.

### 5.2 Cardinality reduction

Dynamic programming for MPS is the essential ingredient of **cardinality reduction**. It was first developed by [25, 1] and then reformulated using the notion of "boundary contraction" [16].

### 5.2.1 Boundary contraction

Let $\text{tr}_{[i,j]} \rho$ denote the partial trace over $\mathcal{H}_{[i,j]}$ of a density matrix $\rho$ on $\mathcal{H}$.

**Definition 5** (boundary contraction). Let $\psi = \sum_{j=1}^B \lambda_j l_j \otimes r_j$ be the Schmidt decomposition of a state $\psi \in \mathcal{H}$ across the cut $i|i+1$. Let $\{|j\rangle\}_{j=1}^B$ be the computational basis of $\mathbf{C}^B$. Let $U_\psi : \mathbf{C}^B \to \mathcal{H}_{[i+1,n]}$ be the isometry specified by $U_\psi|j\rangle = r_j$ such that $U_\psi^{-1}\psi = \sum_{j=1}^B \lambda_j l_j |j\rangle \in \mathcal{H}_{[1,i]} \otimes \mathbf{C}^B$. The boundary contraction $\text{cont}\,\psi$ is a density matrix on $\mathcal{H}_i \otimes \mathbf{C}^B$:

$$\text{cont}\,\psi := U_\psi^{-1} \text{tr}_{[1,i-1]}(|\psi\rangle\langle\psi|)U_\psi. \tag{10}$$

Let $\epsilon_L, \epsilon_R$ be the ground-state energies of $H_L := \sum_{j=1}^{i-1} H_j$ and $H_R := \sum_{j=i+1}^{n-1} H_j$, respectively. Define $H'_L = H_L - \epsilon_L$ and $H'_R = H_R - \epsilon_R$ so that the ground-state energies of $H'_L, H'_R$ are 0.

**Lemma 13.** *Let $\rho$ be a density matrix on $\mathcal{H}_{[1,i]} \otimes \mathbf{C}^B$ and $\psi = \sum_{j=1}^B \lambda_j l_j \otimes r_j$ be the Schmidt decomposition of a state $\psi \in \mathcal{H}$ (across the cut $i|i+1$). The density matrix $\rho' := U_\psi \rho U_\psi^{-1}$ on $\mathcal{H}$ has energy*

$$\text{tr}(\rho' H) \leq \text{tr}(\rho H_L) + \langle \psi, (H_i + H_R)\psi \rangle + \|\text{tr}_{[1,i-1]} \rho - \text{cont}\,\psi\|_1 \left(1 + \max_{r \in \text{span}\{r_j\}} \|H'_R r\|\right). \tag{11}$$

*Proof.* Since $U_\psi$ is a isometry,

$$\begin{aligned}
&\text{tr}(\rho' H) - \text{tr}(\rho H_L) - \langle \psi, (H_i + H_R)\psi \rangle = \text{tr}(\rho' H) - \text{tr}(\rho' H_L) - \langle \psi, (H_i + H_R)\psi \rangle \\
&= \text{tr}(\rho'(H_i + H_R)) - \langle \psi, (H_i + H_R)\psi \rangle = \text{tr}[(\rho' - |\psi\rangle\langle\psi|)(H_i + H_R)] \\
&= \text{tr}[(\rho' - |\psi\rangle\langle\psi|)(H_i + H'_R)] = \text{tr}[\text{tr}_{[1,i-1]}(\rho' - |\psi\rangle\langle\psi|)(H_i + H'_R)] \\
&= \text{tr}[U_\psi^{-1} \text{tr}_{[1,i-1]}(\rho' - |\psi\rangle\langle\psi|) U_\psi U_\psi^{-1}(H_i + H'_R)U_\psi] \\
&= \text{tr}[(\text{tr}_{[1,i-1]}\rho - \text{cont}\,\psi)(U_\psi^{-1} H_i U_\psi + U_\psi^{-1} H'_R U_\psi)] \\
&\leq \|\text{tr}_{[1,i-1]}\rho - \text{cont}\,\psi\|_1 \cdot \|U_\psi^{-1} H_i U_\psi + U_\psi^{-1} H'_R U_\psi\| \\
&\leq \|\text{tr}_{[1,i-1]}\rho - \text{cont}\,\psi\|_1 (1 + \|U_\psi^{-1} H'_R U_\psi\|) \leq \|\text{tr}_{[1,i-1]}\rho - \text{cont}\,\psi\|_1 \left(1 + \max_{r \in \text{span}\{r_j\}} \|H'_R r\|\right) \quad (12)
\end{aligned}$$

$\square$

### 5.2.2 Algorithm

Let $N$ be a $\xi$-net with $\xi = \tilde{\Omega}(\epsilon)$ for the trace norm over the space of boundary contractions of bond dimension $B_{8\sqrt{c}\epsilon^2} = 2^{\tilde{O}(1/\epsilon)}$ so that $|N| = (B/\xi)^{O(B)} = 2^{2^{\tilde{O}(1/\epsilon)}}$. It is straightforward to construct $N$ in time poly $|N| = 2^{2^{\tilde{O}(1/\epsilon)}}$.

===================================================================

**Cardinality reduction** convex program and **bond truncation**

___________________________________________________________________

0. Let the variable $\rho$ be a density matrix on span $S_i^{(1)} \otimes \mathbf{C}^{B_{8\sqrt{c}\epsilon^2}} \subseteq \mathcal{H}_{[1,i]} \otimes \mathbf{C}^{B_{8\sqrt{c}\epsilon^2}}$.
1. For each $X \in N$, solve the convex program:

$$\min \text{tr}(\rho H_L); \text{ s. t. } \|\text{tr}_{[1,i-1]}\rho - X\|_1 \leq \xi, \ \text{tr}\,\rho = 1, \ \rho \geq 0. \tag{13}$$

2. Let $\varphi = \sum_j \varphi_j |j\rangle$ be the eigenvector of the solution $\rho$ with the largest eigenvalue.
3. Let $\varphi' = \sum_j \varphi'_j |j\rangle$ be the state obtained by truncating each bond (in whatever order) of $\varphi$ to $p_2$.
4. $S_i^{(3)}$ consists of the MPS representations of all $\varphi'_j$.

===================================================================

### 5.2.3 Analysis

Let $P_t$ be the projection onto the subspace $\mathcal{H}_R^{\le t}$ spanned by the eigenvectors of $H'_R$ with eigenvalues at most $t$, and $Q_t$ be the projection onto the subspace spanned by the eigenvectors of $H'_L + H'_R$ with eigenvalues at most $t$.

**Lemma 14** (truncation lemma).
$$\|(1-P_t)\Psi_0\| \le \|(1-Q_t)\Psi_0\| \le 100 \cdot 2^{-t/20}. \tag{14}$$

*Proof.* The first inequality is obvious: $P_t \ge Q_t$ as $[H'_L, H'_R] = 0$ and $H'_L \ge 0$. The second inequality was proved in [5]. □

Let $t = O(\log(1/\epsilon))$ so that $100 \cdot 2^{-t/20} \le c\epsilon^6$.

**Lemma 15.** *There exists a state $\psi \in \mathrm{span}\, S_i^{(1)} \otimes \mathcal{H}_R^{\le t}$ of Schmidt rank $B_{8\sqrt{c}\epsilon^2}$ (across the cut $i|i+1$) such that $\langle \psi, H\psi \rangle \le \epsilon_0 + 200 c^{1/4}\epsilon$.*

*Proof.* Let $\phi$ be a witness for $S_i^{(1)}$. Since $S_i^{(1)}$ is an $(i, dp_1 p_3, p_2 p_3, \Delta = c\epsilon^6)$-support set, Lemma 5 implies $|\langle \phi, \Psi_0 \rangle| \ge 1 - c\epsilon^6$. Lemma 7 implies $\langle \phi, H_i \phi \rangle \ge \langle \Psi_0, H_i \Psi_0 \rangle - 2\sqrt{2c}\epsilon^3$. Let $\phi' = P_t\phi/\|P_t\phi\|$ so that $\phi' \in \mathrm{span}\, S_i^{(1)} \otimes \mathcal{H}_R^{\le t}$ by construction. Since the state $\phi$ has energy at most $\epsilon_0 + c\epsilon^7$,

$$\begin{aligned}\langle \Psi_0, (H'_L + H'_R)\Psi_0 \rangle + 4\sqrt{c}\epsilon^3 &\ge \langle \Psi_0, (H'_L + H'_R)\Psi_0 \rangle + 2\sqrt{2c}\epsilon^3 + c\epsilon^7 \ge \langle \phi, (H'_L + H'_R)\phi \rangle \\ &= \langle \phi, P_t(H'_L + H'_R)P_t\phi \rangle + \langle \phi, (1-P_t)(H'_L + H'_R)(1-P_t)\phi \rangle \\ &\ge \langle \phi', (H'_L + H'_R)\phi' \rangle \|P_t\phi\|^2 + t\|(1-P_t)\phi\|^2. \end{aligned} \tag{15}$$

The ground-state energy of $H'_L + H_i + H'_R$ is at most 1 as $H_i \le 1$. Hence, $\langle \Psi_0, (H'_L + H'_R)\Psi_0 \rangle \le 1$ as $H_i \ge 0$. (15) implies

$$t \gg \langle \Psi_0, (H'_L + H'_R)\Psi_0 \rangle + 4\sqrt{c}\epsilon^3 \Rightarrow \langle \phi', (H'_L + H'_R)\phi' \rangle \le \langle \Psi_0, (H'_L + H'_R)\Psi_0 \rangle + 4\sqrt{c}\epsilon^3. \tag{16}$$

Lemmas 7, 14 imply

$$\begin{aligned}|\langle \phi', \Psi_0 \rangle| &\ge |\langle P_t\phi, \Psi_0 \rangle| = |\langle \phi, P_t\Psi_0 \rangle| \ge |\langle \phi, \Psi_0 \rangle| - |\langle \phi, (1-P_t)\Psi_0 \rangle| \ge 1 - c\epsilon^6 - \|(1-P_t)\Psi_0\| \\ &\ge 1 - c\epsilon^6 - 100\cdot 2^{-t/20} \ge 1 - 2c\epsilon^6 \Rightarrow \langle \phi', H_i\phi' \rangle \le \langle \Psi_0, H_i\Psi_0 \rangle + 4\sqrt{c}\epsilon^3. \end{aligned} \tag{17}$$

Summing (16) (17) gives $\langle \phi', H\phi' \rangle \le \langle \Psi_0, H\Psi_0 \rangle + 4\sqrt{c}\epsilon^3 + 4\sqrt{c}\epsilon^3 = \epsilon_0 + 8\sqrt{c}\epsilon^3$. Finally, Lemma 12 implies that the state $\psi := \mathrm{trunc}_{B_{8\sqrt{c}\epsilon^2}}\phi'/\|\mathrm{trunc}_{B_{8\sqrt{c}\epsilon^2}}\phi'\| \in \mathrm{span}\, S_i^{(1)} \otimes \mathcal{H}_R^{\le t}$ has energy $\langle \psi, H\psi \rangle \le \epsilon_0 + 75c^{1/4}\epsilon$. □

**Lemma 16.** *$S_i^{(2)}$ is an $(i, p_1, dp_1p_2p_3^2, \Delta = 1/1000)$-support set, where $S_i^{(2)}$ consists of the MPS representations of all $\varphi_j$.*

*Proof.* Since $N$ is a $\xi$-net, there is an element $X \in N$ such that $\|\mathrm{cont}\,\psi - X\|_1 \le \xi$. Clearly, $\mathrm{tr}(\rho H_L) \le \langle \psi, H_L\psi \rangle$ as $U_\psi^{-1}|\psi\rangle\langle\psi|U_\psi$ is a feasible solution to the convex program (13). Let $\sigma = U_\psi \rho U_\psi^{-1}$, and set $\xi = \tilde{\Omega}(\epsilon)$ such that $2\xi(1+t) \le \epsilon/4000$. Lemma 13 implies

$$\begin{aligned}\mathrm{tr}(\sigma H) &\le \mathrm{tr}(\rho H_L) + \langle \psi, (H_i + H_R)\psi \rangle + \|\mathrm{tr}_{[1,i-1]}\rho - \mathrm{cont}\,\psi\|_1 \left(1 + \max_{r \in \mathrm{span}\{r_j\}} \|H'_R r\|\right) \\ &\le \langle \psi, (H_L + H_i + H_R)\psi \rangle + 2\xi\left(1 + \max_{r \in \mathcal{H}_R^{\le t}} \|H'_R r\|\right) \le \langle \psi, H\psi \rangle + 2\xi(1+t) \\ &\le \epsilon_0 + 200c^{1/4}\epsilon + \epsilon/4000 \le \epsilon_0 + \epsilon/2000 \end{aligned} \tag{18}$$

for sufficiently small constant $c$. We observe that

(1) there exists at least an eigenstate of $\sigma$ with energy at most $\epsilon_0 + \epsilon/1000$;
(2) there is at most one such eigenstate as Lemma 5 implies that such an eigenstate is close to $\Psi_0$;
(3) this eigenstate (denoted by $\Phi$) has the largest eigenvalue due to Markov's inequality in probability theory;
(4) $\Phi = U_\psi \varphi$ is a witness for $S_i^{(2)}$ as an $(i, p_1, dp_1 p_2 p_3^2, \Delta = 1/1000)$-support set with $p_1 = B_{8\sqrt{c}\epsilon^2}|N| = 2^{2^{\tilde{O}(1/\epsilon)}}$. $\square$

## 5.3 Bond truncation

The analysis of **bond truncation** follows immediately from Lemmas 1, 9, and is (almost) identical to that in [16].

**Lemma 17.** $S_i^{(3)}$ is an $(i, p_1, p_2, \delta = 1/20)$-support set.

*Proof.* Since $\Phi$ is a witness for $S_i^{(2)}$ with energy at most $\epsilon_0 + \epsilon/1000$, Lemma 5 implies $|\langle \Phi, \Psi_0 \rangle| \geq 999/1000$. Lemma 1 implies the existence of an MPS $\Psi$ of bond dimension $2^{\tilde{O}(\epsilon^{-1/4} \log^{3/4} n)}$ such that $|\langle \Psi, \Psi_0 \rangle| \geq 999/1000$. Lemma 6 implies $|\langle \Psi, \Phi \rangle| \geq 249/250$. Let $\Phi'$ be the state obtained by truncating each bond to the left of the cut $i|i+1$ (in the same order as each bond of $\varphi$ is truncated) of $\Phi$ to $p_2 = 1000n2^{\tilde{O}(\epsilon^{-1/4} \log^{3/4} n)}$. Lemma 9 implies $|\langle \Psi, \Phi' \rangle| \geq 199/200$. Lemma 6 implies $|\langle \Psi_0, \Phi' \rangle| \geq 49/50$. Hence $\Phi' = U_\psi \varphi'$ is a witness for $S_i^{(3)}$ as an $(i, p_1, p_2, \delta = 1/20)$-support set. $\square$

## 5.4 Error reduction

To reduce the energy of the witness $\Phi'$, we efficiently construct an MPO that approximately projects $\Phi'$ onto the ground state $\Psi_0$. This construction and its variants were used extensively by Hastings to prove several well-known results (e.g., exponential decay of correlations [11, 14] and the 1D area law for entanglement [12]) for the ground state of gapped local Hamiltonians. It applies to general 1D gapped systems and does not involve randomness. It uses the Fourier transform and the Lieb-Robinson bound [18, 19, 14].

Assume for the moment that we have an estimate $\epsilon_0'$ of the ground-state energy $\epsilon_0$ of $H$ in the sense that $|\epsilon_0 - \epsilon_0'| \leq \epsilon/\sqrt{q} \leq \epsilon/2$, where $q = 4\log(1/\eta') + 24$ with $\eta'$ to be specified later. Let

$$A := e^{-\frac{q(H-\epsilon_0')^2}{2\epsilon^2}} = \frac{\epsilon}{\sqrt{2\pi q}} \int_{-\infty}^{+\infty} e^{-\frac{\epsilon^2 t^2}{2q} - i(H-\epsilon_0')t} \mathrm{d}t. \quad (19)$$

**Lemma 18.** $|\langle \Phi', \Psi_0 \rangle| \geq 19/20$ implies $\langle \phi, H\phi \rangle \leq \epsilon_0 + \eta'\epsilon/100$ for $\phi = A\Phi'/\|A\Phi'\|$.

*Proof.* Let $\{\Psi_j\}_{j=0}^{d^n-1}$ be the eigenvectors of $H$ with the corresponding eigenvalues $\{\epsilon_j\}_{j=0}^{d^n-1}$ in ascending order. The state $\Phi'$ can be decomposed as $\Phi' = \sum_{j=0}^{d^n-1} c_j \Psi_j$ with $|c_0| \geq 19/20$.

$$\|A\Phi'\| \geq |c_0| \cdot \|A\Psi_0\| \geq 19e^{-\frac{q(\epsilon_0 - \epsilon_0')^2}{2\epsilon^2}}/20 \geq 19/(20\sqrt{e}) \geq 1/2. \quad (20)$$

Since $\epsilon_j - \epsilon_0 \leq 2(\epsilon_j - \epsilon_0')$ for $j \geq 1$,

$$\langle \phi, (H-\epsilon_0)\phi \rangle = \langle A\Phi', (H-\epsilon_0)A\Phi' \rangle/\|A\Phi'\|^2 \leq 4 \sum_{j=1}^{d^n-1} (\epsilon_j - \epsilon_0)|c_j|^2 e^{-q(\epsilon_j - \epsilon_0')^2/\epsilon^2}$$

$$\leq 8 \sum_{j=1}^{d^n-1} (\epsilon_j - \epsilon_0')|c_j|^2 e^{-q(\epsilon_j - \epsilon_0')^2/\epsilon^2} \leq 8 \max_{x \geq \epsilon/2}\{xe^{-qx^2/\epsilon^2}\} \sum_{j=1}^{d^n-1} |c_j|^2 \leq 4\epsilon e^{-q/4} \leq \eta'\epsilon/100. \quad (21)$$

**Lemma 19.** *An MPO $K$ of bond dimension $D = n^{O(1+\sqrt{\log(1/\eta')/\log n}/\epsilon)}$ can be efficiently constructed such that $\|K - A\| \leq \eta'\epsilon/(1000n) =: \eta''$.*

*Proof.* Following [13], we truncate and discretize the integral in (19):

$$A = \frac{\epsilon}{\sqrt{2\pi q}} \int_{-\infty}^{+\infty} e^{-\frac{\epsilon^2 t^2}{2q} - i(H-\epsilon_0')t} \mathrm{d}t \approx \frac{\epsilon}{\sqrt{2\pi q}} \int_{-T}^{T} e^{-\frac{\epsilon^2 t^2}{2q} - i(H-\epsilon_0')t} \mathrm{d}t$$

$$\approx \frac{\epsilon\tau}{\sqrt{2\pi q}} \sum_{j=-T/\tau}^{T/\tau} e^{-\frac{\epsilon^2 \tau^2 j^2}{2q} - i(H-\epsilon_0')\tau j} \approx \frac{\epsilon\tau}{\sqrt{2\pi q}} \sum_{j=-T/\tau}^{T/\tau} e^{-\frac{\epsilon^2 \tau^2 j^2}{2q}} K_j =: K. \quad (22)$$

The truncation error is of order $e^{-\epsilon^2 T^2/(2q)} \leq \eta''/3$ for $T = O(\sqrt{q\log(1/\eta'')}/\epsilon)$. The discretization error is of order $\epsilon\tau\|H\|T/\sqrt{2\pi q} \leq \eta''/3$ for $\tau = O(\eta''\sqrt{q}/(\epsilon\|H\|T)) = O(\eta''/(n\sqrt{\log(1/\eta'')}))$. As a consequence of the Lieb-Robinson bound [18], each propagator $e^{-i(H-\epsilon_0')\tau j}$ can be approximated to precision $\eta''/3$ by an MPO $K_j$ of bond dimension [20] $2^{O(\tau j)}\mathrm{poly}(n/\eta'') = 2^{O(T)}\mathrm{poly}(n/\eta'') = 2^{O(\sqrt{q\log(1/\eta'')}/\epsilon)}\mathrm{poly}(n/\eta'')$. The number of terms is $2T/\tau + 1 = O(n\sqrt{q}\log(1/\eta'')/(\eta''\epsilon))$. Therefore, the bond dimension of the MPO $K$ is $D = n^{O(1+\sqrt{\log(1/\eta')/\log n}/\epsilon)}$. □

Lemmas 18, 19 imply $\langle\phi', H\phi'\rangle \leq \epsilon_0 + \eta'\epsilon/50$ for $\phi' = K\Phi'/\|K\Phi'\|$. Since $K$ is an MPO of bond dimension $D$, it can be decomposed as $K = \sum_{j=1}^{D} L_j \otimes R_j$, where $L_j$ and $R_j$ are MPO of bond dimension $D$ on $\mathcal{H}_{[1,i]}$ and on $\mathcal{H}_{[i+1,n]}$, respectively. It is easy to see that $S_i := \{L_j\psi : j = 1,2,\ldots,D, \forall \psi \in S_i^{(3)}\}$ is an $(i, p_1 D, p_2 D, \Delta = \eta')$-support set. Hence $p_3 = n^{O(1+\sqrt{\log(1/\eta')/\log n}/\epsilon)} = n^{O(1)}$ by setting $\eta' = c\epsilon^6$.

## 5.5 The final step

By minimizing the energy over the subspace span $S_n$, we obtain an MPS $\Psi^{(0)}$ of bond dimension $n^{O(1)}$ such that $|\langle\Psi^{(0)}, \Psi_0\rangle| \geq 1 - c\epsilon^6 \geq 1 - c$. Given $\eta = n^{-O(1)}$, we inductively construct a sequence of MPS $\{\Psi^{(j)}\}_{j=1}^{\log_2(c/\eta)}$ of bond dimension $p_4$ such that $|\langle\Psi^{(j)}, \Psi_0\rangle| \geq 1 - c/2^j$, where $p_4 = n^{1+o(1)}/\eta$ is a fixed polynomial in $n$.

Let $\eta' = 1$. Based on a minor modification of the proof of Lemma 19, we efficiently construct an MPO $K$ of bond dimension $D = n^{O(1)}$ such that $\|K - A\| \leq \eta/1000$. Given a desirable $\Psi^{(j-1)}$, it is easy to see $|\langle\Phi^{(j)}, \Psi_0\rangle| \geq 1 - c/2^{j+4}$ for $\Phi^{(j)} = K\Psi^{(j-1)}/\|K\Psi^{(j-1)}\|$. We obtain $\Psi^{(j)}$ by truncating each bond (in whatever order) of $\Phi^{(j)}$ to $p_4$. A minor modification of the proof of Lemma 17 implies $|\langle\Psi^{(j)}, \Psi_0\rangle| \geq 1 - c/2^j$.

We briefly comment on the assumption that we have an estimate $\epsilon_0'$ of the ground-state energy $\epsilon_0$ of $H$ in the sense that $|\epsilon_0 - \epsilon_0'| \leq \epsilon/\sqrt{q}$. Since $0 \leq \epsilon_0 \leq n$, we run the whole algorithm with $\epsilon_0' = j\epsilon/\sqrt{q}$ and obtain a candidate MPS solution for each $j = 0, 1, \ldots, \lceil n\sqrt{q}/\epsilon\rceil$. The candidate MPS with the lowest energy is identified as the final output.

## 6 Frustration-free systems

To prove Theorem 2, the only step that needs to be modified is **error reduction**. In particular, we construct a simpler and more efficient MPO to reduce the energy of $\Phi'$ in frustration-free systems using a slightly improved (and simplified) version [4] of the detectability lemma [2]. Assume without loss of generality that each term $H_i$ in the frustration-free 1D gapped Hamiltonian $H = \sum_{i=1}^{n-1} H_i$ is a projector, i.e., $H_i^2 = H_i$. Then $1 - H_i$ is the projection onto the ground-state space of $H_i$.

**Lemma 20** (detectability lemma [4]). *Define $A = \prod_i (1 - H_i)$, where the product is taken in an arbitrary order of $(1 - H_i)$'s. Then for any state $\Psi_1'$,*

$$\langle \Psi_1', A^\dagger H A \Psi_1' \rangle \leq 4(1 - \|A\Psi_1'\|^2). \tag{23}$$

*This implies that for any state $\Psi_1 \perp \Psi_0$,*

$$\|A\Psi_1\| \leq 1/\sqrt{1 + \epsilon/4} =: f(\epsilon) = 1 - \Omega(\epsilon). \tag{24}$$

Recall from the proof of Lemma 17 that $\Phi'$ is a witness for $S_i^{(3)}$ with $|\langle \Phi', \Psi_0 \rangle| \geq 19/20$.

**Lemma 21.** *An MPO $K$ of bond dimension $D = 1/(\eta'\epsilon)^{O(1/\epsilon)}$ can be efficiently constructed such that $\langle \phi', H\phi' \rangle \leq \eta'\epsilon$ for $\phi' = K\Phi'/\|K\Phi'\|$.*

*Proof.* Define $K' = A^l$ and $K = A^{l+1}$. Clearly, $K'\Psi_0 = \Psi_0$ and $\langle \Psi_0, K'\Psi_1 \rangle = 0$ for any $\Psi_1 \perp \Psi_0$. Furthermore, (24) implies $\|K'\Psi_1\| \leq e^{-\Omega(l\epsilon)}$. Let $\Psi_1' = K'\Phi'/\|K'\Phi'\|$ such that $\phi' = A\Psi_1'/\|A\Psi_1'\|$. (23) implies

$$\|A\Psi_1'\| \geq |\langle \Psi_1', \Psi_0 \rangle| \geq 1 - e^{-\Omega(l\epsilon)} \Rightarrow \langle \phi', H\phi' \rangle \leq 4(1/\|A\Psi_1'\|^2 - 1) \leq e^{-\Omega(l\epsilon)} = \eta'\epsilon \tag{25}$$

for $l = O(\log(1/(\eta'\epsilon))/\epsilon)$. Since $A$ is an MPO of bond dimension $d^2$, the bond dimension of $K$ is $D = O(d^{2l+2}) = 1/(\eta'\epsilon)^{O(1/\epsilon)}$. □

Hence $p_3 = 2^{\tilde{O}(1/\epsilon)}$ by setting $\eta' = c\epsilon^6$. Furthermore, the bond dimension of $K$ is $D = n^{o(1)}$ for $\eta' = n^{-o(1)}$. The exponent $\omega + 2 + o(1)$ in Theorem 2 comes from the observation that all convex programs are of subpolynomial size. For each $i = 1, 2, \ldots, n$, the construction of these convex programs requires contracting MPS (whose bond dimension is $n^{1+o(1)}$) in $S_i$.

## 7 Degenerate ground states

Previously we assumed a unique ground state. We now extend all results to the case of degenerate ground states. Shortly after version 3 of the present preprint appeared on arXiv (version 1 does not contain any results for degenerate ground states), a very different approach of extending results from unique to degenerate ground states was given independently by Chubb and Flammia [6]. See "note added in preparation" in [6] for a detailed discussion of the differences between their approach and the one presented here.

### 7.1 Main results, overview, and preliminaries

Suppose the ground states of a 1D Hamiltonian $H = \sum_{i=1}^{n-1} H_i$ are two-fold exactly degenerate (for ease of presentation), and there is a constant energy gap $\epsilon$. It should be clear that a minor modification of the proof works for any constant-fold degeneracy and leads to the same results in the presence of an exponentially small $2^{-\Omega(n)}$ splitting of the degeneracy (as is typically observed in physical systems). The goal is to find efficient MPS approximations to a set of basis vectors of the ground-state space $G$ of $H$. The existence of an efficient MPS approximation to any ground state $\Psi_0 \in G$ of $H$ is a by-product of the proof of the area law for entanglement [15]. Let $\eta = n^{-O(1)}$.

**Lemma 22** ([15]). *For any ground state $\Psi_0 \in G$ of $H$, there exists an MPS $\Psi$ of bond dimension $2^{\tilde{O}(1/\epsilon + \epsilon^{-1/4} \log^{3/4}(n/\eta))}$ such that $|\langle \Psi, \Psi_0 \rangle| \geq 1 - \eta$.*

The main result of the present section is

**Theorem 3.** *In 1D gapped systems with two-fold ground-state degeneracy there is an $n^{O(1)}$-time algorithm that outputs two orthogonal MPS $\Psi, \Psi'$ such that $|\langle \Psi, \Psi_0 \rangle| \geq 1 - \eta$ and $|\langle \Psi', \Psi_1 \rangle| \geq 1 - \eta$, where $\Psi_0, \Psi_1 \in G$ are two ground states of $H$.*

Theorem 2 can be extended to the case of frustration-free degenerate ground states in the same way.

**Definition 6** (support set). $S \subseteq \mathcal{H}_{[1,i]}$ is an $(i, s, b, \delta$ or $\Delta)$-support set if there exist two orthogonal states $\psi_0, \psi_1 \in \mathcal{H}$ (each of which is called a witness for $S$) such that
(i) the reduced density matrices of both $\psi_0$ and $\psi_1$ on $\mathcal{H}_{[1,i]}$ are supported on span $S$ (hence, the reduced density matrix of any state $\psi \in \mathrm{span}\{\psi_0, \psi_1\}$ on $\mathcal{H}_{[1,i]}$ is supported on span $S$);
(ii) $|S| \leq s$;
(iii) all elements in $S$ are MPS of bond dimension at most $b$;
(iv) there are two ground states $\Psi_0, \Psi_1 \in G$ of $H$ such that $|\langle \psi_0, \Psi_0 \rangle| \geq 1 - \delta$ and $|\langle \psi_1, \Psi_1 \rangle| \geq 1 - \delta$;
or (iv) $\langle \psi_0, H\psi_0 \rangle \leq \epsilon_0 + \Delta\epsilon$ and $\langle \psi_1, H\psi_1 \rangle \leq \epsilon_0 + \Delta\epsilon$ (depending on the context either $\delta$ or $\Delta$ is used as the precision parameter).

There are no major changes in the outline (Section 3) of the algorithm. Support sets $S_i$'s are iteratively constructed for $i = 1, 2, \ldots, n$, and Table 1 illustrates the evolution of the parameters at every step in each iteration. A few lemmas in Section 4 should be modified, and a new lemma is added.

**Lemma 23.** $\langle \psi, H\psi \rangle \leq \epsilon_0 + \eta\epsilon$ *implies the existence of a ground state $\Psi_0 \in G$ such that $|\langle \psi, \Psi_0 \rangle|^2 \geq 1 - \eta$.*

**Lemma 24** ([15]). $\langle \Psi'_0, \Psi_0 \rangle \geq 1 - \eta$ *for any ground state $\Psi_0 \in G$ of $H$ and $\Psi'_0 = \mathrm{trunc}_{B_\eta} \Psi_0 / \| \mathrm{trunc}_{B_\eta} \Psi_0 \|$, where $B_\eta = 2^{\tilde{O}(1/\epsilon + \epsilon^{-1/4} \log^{3/4}(1/\eta))}$.*

**Lemma 25.** $\langle \psi_0, H\psi_0 \rangle \leq \epsilon_0 + \Delta\epsilon$ *and $\langle \psi_1, H\psi_1 \rangle \leq \epsilon_0 + \Delta\epsilon$ imply $\langle \psi, H\psi \rangle \leq \epsilon_0 + 2\Delta\epsilon/(1 - |\langle \psi_0, \psi_1 \rangle|)$ for any state $\psi \in \mathrm{span}\{\psi_0, \psi_1\}$.*

*Proof.* Any state $\psi = \alpha\psi_0 + \beta\psi_1 \in \mathrm{span}\{\psi_0, \psi_1\}$ has energy at most $\epsilon_0 + 2\Delta\epsilon/(1 - |\langle \psi_0, \psi_1 \rangle|)$ as

$$1 = \|\psi\|^2 \geq |\alpha|^2 + |\beta|^2 - 2|\alpha| \cdot |\beta| \cdot |\langle \psi_0, \psi_1 \rangle| \Rightarrow |\alpha|^2 + |\beta|^2 \leq 1/(1 - |\langle \psi_0, \psi_1 \rangle|)$$
$$\Rightarrow \langle \psi, (H - \epsilon_0)\psi \rangle \leq |\alpha|^2 \langle \psi_0, (H - \epsilon_0)\psi_0 \rangle + |\beta|^2 \langle \psi_1, (H - \epsilon_0)\psi_1 \rangle + 2|\alpha| \cdot |\beta| \cdot |\langle \psi_0, (H - \epsilon_0)\psi_1 \rangle|$$
$$\leq (|\alpha|^2 + |\beta|^2)\Delta\epsilon + 2|\alpha| \cdot |\beta| \sqrt{\langle \psi_0, (H - \epsilon_0)\psi_0 \rangle \cdot \langle \psi_1, (H - \epsilon_0)\psi_1 \rangle} \leq 2(|\alpha|^2 + |\beta|^2)\Delta\epsilon. \quad (26)$$

□

## 7.2 Algorithm and analysis

In the $i$th iteration the algorithm in Theorem 3 constructs an $(i, p_1 p_3, p_2 p_3, \Delta = c\epsilon^6)$-support set $S_i$ from an $(i-1, p_1 p_3, p_2 p_3, \Delta = c\epsilon^6)$-support set $S_{i-1}$ returned in the $(i-1)$th iteration. **Extension** is trivial, and we obtain an $(i, dp_1 p_3, p_2 p_3, \Delta = c\epsilon^6)$-support set $S_i^{(1)}$. Let $N$ be a $\xi$-net over the space of boundary contractions of bond dimension $2B_{8\sqrt{c}\epsilon^2}$.

===============================================================

**Cardinality reduction** convex program and **bond truncation**

---

0. Let the variables $\rho_0, \rho_1$ be density matrices on span $S_i^{(1)} \otimes \mathbf{C}^{2B_{8\sqrt{c}\epsilon^2}} \subseteq \mathcal{H}_{[1,i]} \otimes \mathbf{C}^{2B_{8\sqrt{c}\epsilon^2}}$.
1. For each $X \in N$, solve the convex program:

$$\min \mathrm{tr}(\rho_0 H_L); \text{ s. t. } \| \mathrm{tr}_{[1,i-1]} \rho_0 - X \|_1 \leq \xi, \ \mathrm{tr}\, \rho_0 = 1, \ \rho_0 \geq 0. \quad (27)$$

2. Let $\varphi_0 = \sum_j \varphi_{0,j} |j\rangle$ be the eigenvector of the solution $\rho_0$ with the largest eigenvalue.
3. Solve the convex program:

$$\min \operatorname{tr}(\rho_1 H_L); \text{ s. t. } \|\operatorname{tr}_{[1,i-1]} \rho_1 - X\|_1 \leq \xi, \ \operatorname{tr} \rho_1 = 1, \ \rho_1 \geq 0, \ \langle \varphi_0, \rho_1 \varphi_0 \rangle \leq 1/4. \tag{28}$$

4. Let $\varphi_1 = \sum_j \varphi_{1,j} |j\rangle$ be the eigenvector of the solution $\rho_1$ with the largest eigenvalue.
5. Let $\varphi'_0 = \sum_j \varphi'_{0,j} |j\rangle$ and $\varphi'_1 = \sum_j \varphi'_{1,j} |j\rangle$ be the states obtained by truncating each bond (in whatever order) of $\varphi_0$ and $\varphi_1$ to $p_2$, respectively.
6. $S_i^{(3)}$ consists of the MPS representations of all $\varphi'_{0,j}, \varphi'_{1,j}$.
================================================================

**Lemma 26** ([15]). *For any ground state $\Psi_0 \in G$ of $H$,*

$$\|(1 - P_t)\Psi_0\| \leq 100 \cdot 2^{-t/20}. \tag{29}$$

**Lemma 27.** *There exist two states $\psi_0, \psi_1 \in \operatorname{span} S_i^{(1)} \otimes \mathcal{H}_R^{\leq t}$ of Schmidt rank $B_{8\sqrt{c}\epsilon^2}$ (across the cut $i|i+1$) such that $\langle \psi, H\psi \rangle \leq \epsilon_0 + 200 c^{1/4} \epsilon$ for any state $\psi \in \operatorname{span}\{\psi_0, \psi_1\}$.*

*Proof.* Let $\phi_0 \perp \phi_1$ be two orthogonal witnesses for $S_i^{(1)}$. Define $\phi'_0 = P_t \phi_0 / \|P_t \phi_0\|$. The proof of Lemma 15 implies that the state $\psi_0 := \operatorname{trunc}_{B_{8\sqrt{c}\epsilon^2}} \phi'_0 / \|\operatorname{trunc}_{B_{8\sqrt{c}\epsilon^2}} \phi'_0\|$ has energy $\langle \psi_0, H\psi_0 \rangle \leq \epsilon_0 + 75 c^{1/4} \epsilon$. Since $\phi_0$ is a low-energy state, Markov's inequality in probability theory implies $|\langle \phi_0, \phi'_0 \rangle| \geq 999/1000$ for sufficiently large $t$. The proof of Lemma 12 implies $|\langle \psi_0, \phi'_0 \rangle| \geq 999/1000$ for sufficiently small $c$. Lemma 6 implies $|\langle \psi_0, \phi_0 \rangle| \geq 249/250$. Similarly, we obtain another state $\psi_1$ such that $\langle \psi_1, H\psi_1 \rangle \leq \epsilon_0 + 75 c^{1/4} \epsilon$ and $|\langle \psi_1, \phi_1 \rangle| \geq 249/250$. It is easy to see $|\langle \psi_0, \psi_1 \rangle| \leq 1/5$, and Lemma 25 implies that $\langle \psi, H\psi \rangle \leq \epsilon_0 + 200 c^{1/4} \epsilon$ for any state $\psi \in \operatorname{span}\{\psi_0, \psi_1\}$. $\square$

**Lemma 28.** *For any state $\psi \in \operatorname{span}\{\psi_0, \psi_1\}$, let $X \in N$ be the element that is closest to $\operatorname{cont} \psi$. Then, at least one of the following must hold:*
*(i) $\Phi_0 = U_\psi \varphi_0$ has energy at most $\epsilon_0 + \epsilon/20000$ and satisfies $|\langle \psi, \Phi_0 \rangle| \geq 1/2$;*
*(ii) $\Phi_0 = U_\psi \varphi_0$ and $\Phi_1 = U_\psi \varphi_1$ have energies at most $\epsilon_0 + \epsilon/20000$ and satisfy $|\langle \Phi_0, \Phi_1 \rangle| \leq \sqrt{3}/2$.*

*Proof.* The proof of Lemma 16 implies that $\sigma_0 = U_\psi \rho_0 U_\psi^{-1}$ has energy $\operatorname{tr}(\sigma_0 H) \leq \epsilon_0 + \epsilon/60000$ for sufficiently small $\xi = \tilde{\Omega}(\epsilon)$ and sufficiently small constant $c$. We observe that (1) there exists at least an eigenstate of $\sigma_0$ with energy at most $\epsilon_0 + \epsilon/20000$; (2) there are at most two such eigenstates as Lemma 23 implies that such eigenstates are close to the ground-state space $G$; (3) one of these eigenstates has the largest eigenvalue (at least $1/3$) due to Markov's inequality in probability theory; (4) this eigenstate is $\Phi_0 = U_\psi \varphi_0$. Therefore, (i) holds if $|\langle \psi, \Phi_0 \rangle| \geq 1/2$. Otherwise, $U_\psi^{-1} |\psi\rangle\langle\psi| U_\psi$ is a feasible solution to the second convex program (28), and the proof of Lemma 16 implies that $\sigma_1 = U_\psi \rho_1 U_\psi^{-1}$ has energy $\operatorname{tr}(\sigma_1 H) \leq \epsilon_0 + \epsilon/60000$. Let $\Phi_1 = U_\psi \varphi_1$ be the eigenvector of $\sigma_1$ with the largest eigenvalue $\lambda$. Similarly, we observe $\lambda \geq 1/3$ and $\langle \Phi_1, H\Phi_1 \rangle \leq \epsilon_0 + \epsilon/20000$. Therefore, (ii) holds because $\langle \Phi_0, \sigma_1 \Phi_0 \rangle \leq 1/4$ implies $|\langle \Phi_0, \Phi_1 \rangle| \leq \sqrt{3}/2$. $\square$

*Remark.* Lemma 25 implies that $S_i^{(2)}$ is an $(i, p_1, dp_1 p_2 p_3^2, \Delta = 1/1000)$-support set with $p_1 = O(B_{8\sqrt{c}\epsilon^2} |N|) = 2^{2^{\tilde{O}(1/\epsilon)}}$, where $S_i^{(2)}$ consists of the MPS representations of all $\varphi_{0,j}, \varphi_{1,j}$.

**Lemma 29.** $S_i^{(3)}$ *is an $(i, p_1, p_2, \delta = 1/20)$-support set.*

*Proof.* $\Phi_k$ and $\varphi_k$ for $k = 0, 1$ in Lemma 28 are functions of $\psi \in \operatorname{span}\{\psi_0, \psi_1\}$. We make this explicit by using the notations $\Phi_k(\psi), \varphi_k(\psi)$. Lemma 28 implies the existence of $\psi, \psi' \in \operatorname{span}\{\psi_0, \psi_1\}$ and $k, k' \in \{0, 1\}$ such that $\Phi_k(\psi)$ and $\Phi_{k'}(\psi')$ have energies at most $\epsilon_0 + \epsilon/20000$ and satisfy $|\langle\Phi_k(\psi), \Phi_{k'}(\psi')\rangle| \leq \sqrt{3}/2$. Let $\varphi'_k(\psi)$ be the state obtained by truncating each bond (in whatever order) of $\varphi_k(\psi)$ to $p_2 = 10000n2^{\tilde{O}(\epsilon^{-1/4}\log^{3/4}n)}$, and $\Phi'_k(\psi)$ be the state obtained by truncating each bond to the left of the cut $i|i+1$ (in the same order) of $\Phi_k(\psi)$ to $p_2$. A minor modification of the proof of Lemma 17 implies the existence of a ground state $\Psi_0 \in G$ such that $|\langle\Phi'_k(\psi), \Psi_0\rangle| \geq 999/1000$ (i.e., $\Phi'_k(\psi) = U_\psi \varphi'_k(\psi)$ is a witness for $S_i^{(3)}$) and $|\langle\Phi_k(\psi), \Phi'_k(\psi)\rangle| \geq 499/500$. Similarly, $\Phi'_{k'}(\psi') = U_{\psi'}\varphi'_k(\psi')$ is another witness with $|\langle\Phi_{k'}(\psi'), \Phi'_{k'}(\psi')\rangle| \geq 499/500$. We obtain two orthogonal witnesses $\Phi'_0, \Phi'_1 \in \mathrm{span}\{\Phi'_k(\psi), \Phi'_{k'}(\psi')\}$ for $S_i^{(3)}$ as an $(i, p_1, p_2, \delta = 1/20)$-support set. □

Lemmas 18, 19 imply two witnesses $\phi'_0 = K\Phi'_0/\|K\Phi'_0\|$ and $\phi'_1 = K\Phi'_1/\|K\Phi'_1\|$ for $S_i$ with energies $\langle\phi'_0, H\phi'_0\rangle \leq \epsilon_0 + \eta'\epsilon/50$ and $\langle\phi'_1, H\phi'_1\rangle \leq \epsilon_0 + \eta'\epsilon/50$. It is easy to see $|\langle\phi'_0, \Phi'_0\rangle| \geq \sqrt{3}/2$ and $|\langle\phi'_1, \Phi'_1\rangle| \geq \sqrt{3}/2$. Hence $|\langle\phi'_0, \phi'_1\rangle| \leq \sqrt{3}/2$, and Lemma 25 implies that any state in $\mathrm{span}\{\phi'_0, \phi'_1\}$ has energy at most $\epsilon_0 + \eta'\epsilon/3$. Therefore, $S_i$ is a support set with $\Delta = \eta' := c\epsilon^6$. Similarly, it is straightforward to extend the "final step" (Section 5.5) to the case of degenerate ground states.